\documentclass{PoS}

\title{Planck scale from broken local conformal invariance in Weyl geometry}

\ShortTitle{Symmetry breakdown of conformal symmetry}

\author{\speaker{Ichiro Oda}\\
        Department of Physics, Faculty of Science, University of the 
           Ryukyus,\\
           Nishihara, Okinawa 903-0213, Japan   \\
        E-mail: \email{ioda@sci.u-ryukyu.ac.jp}}

\abstract{We show that in a quadratic gravity based on Weyl's conformal geometry, the Planck mass scale can be 
generated from quantum effects of the gravitational field and the Weyl gauge field via the Coleman-Weinberg 
mechanism where a local scale symmetry is broken. At the same time, the Weyl gauge field acquires a mass 
less than the Planck mass by absorbing the scalar graviton. The shape of the effective potential is almost flat 
owing to a gravitational character and high symmetries, so our model would provide for an attractive model 
for the inflationary universe. We also present a toy model showing spontaneous symmetry breakdown of 
a global scale symmetry by moving from the Jordan frame to the Einstein one, and point out its problems.}

\FullConference{Corfu Summer Institute 2019 "School and Workshops on Elementary Particle Physics and Gravity" (CORFU2019)\\
		31 August - 25 September 2019\\
		Corfu, Greece}

\begin{document}

\section{Introduction}

One of the most important problems in modern particle physics is to understand the origin of not only the mass 
of elementary particles but also different mass scales existing in nature. This understanding is also important for 
attacking unsolved problems such as the origin of the Higgs potential, the gauge hierarchy problem 
and the cosmological constant problem etc. 

In order to understand the origin of the mass and various mass scales, it is natural to start with a theory without
intrinsic mass scales and consider how the mass is generated from a massless world via dynamical symmetry breaking
mechanism. At this point, let us recall that the mass, or equivalently, the energy, couples to a gravitational field through 
the energy-momentum tensor in a universal manner, so we are forced to take a gravity into consideration for undertanding 
the origin of the mass.  Moreover, it is worthwhile to point out that there naturally appears a local or global scale 
symmetry in a theory having no intrinsic mass scales.  However, since as stressed in \cite{Oda1}, global symmetries 
are in general against the spirit of general relativity (GR) owing to no-hair theorem of black holes \cite{MTW}, one should 
work with a gravitational theory which is invariant under not global but local scale transformation as well as 
the general coordinate transformation at very short distances.\footnote{In this article, we call a global scale symmetry
simply $\textit{scale symmetry}$ while we refer to a local scale transformation as $\textit{conformal symmetry}$ 
by following the terminology of the textbook \cite{Fujii}.}  

In this article, we therefore would like to consider a problem of how we could generate the Planck mass scale, beyond
which the concept of the space-time does not make sense, by beginning with conformally invariant gravitational
theories. From the success of the standard model (SM) of elementary particles, we are confident of the existence
of at least two mass scales, those are, the electroweak scale around $10^2 GeV$ by the Higgs condensation and 
the QCD scale around $10^2 MeV$ by chiral symmetry breaking. These mass scales should be generated via dynamical 
symmetry breakings as well after the Planck mass scale is generated.  

By the way, which conformally invariant gravitational theory is most interesting from the geometrical viewpoint?
We think that it is a Weyl conformal gravity. About one hundred years ago, shortly after the advent of GR by Einstein, 
a conformally invariant extension of GR was proposed by Weyl on the basis of his conformal geometry, what we call, 
the Weyl geometry \cite{Weyl, O'Raifeartaigh}.\footnote{See Ref. \cite{Scholz} for historical review on the Weyl geometry
and various related works \cite{Omote}-\cite{Oda5}.}  
The Weyl geometry is defined as a geometry equipped with a real symmetric metric tensor $g_{\mu\nu}$ as in GR and 
a symmetric connection $\tilde \Gamma^\lambda _{\mu\nu}$, which is related to the Christoffel symbol
$\Gamma^\lambda _{\mu\nu}$ by the relation Eq. (\ref{W-connection}) as seen shortly. It turns out that the Weyl 
geometry reduces to the Riemann geometry when the Weyl gauge field $S_\mu$ is vanishing, or more precisely speaking, 
$S_\mu$ is a gradient, i.e., pure gauge.

In geometrical terms, the Weyl geometry critically differs from the Riemann one in that only angles, but not lengths,
are preserved under parallel transport. To put differently, parallel displacement of a vector field changes its length
in such a way that the very notion of lengths becomes path-dependent. For instance, one can envisage a space traveller, 
who travels to a distant star and then returns to the earth, being surprised to know not only that people in the earth 
have aged much rather than him as predicted by GR in the Riemann geometry but also that the clock on the rocket 
runs at a different rate from those in the earth as understood by Weyl conformal gravity in the Weyl geometry, 
what is called, "the second clock problem" \cite{Penrose}.  Based on this very striking geometry, Weyl has attempted to 
geometrize the electromagnetic theory in the space-time geometry, but his attempt has failed since it turned out later  
that the electromagnetic theory is based on a compact $U(1)$ gauge group whereas the Weyl geometry deals with 
conformal symmetry which is essentially a non-compact Abelian group \cite{O'Raifeartaigh}. Nevertheless, 
it seems that a Weyl quadratic gravity has recently revived as a theory predicting an elementary particle constituting 
dark matter, which is the Weyl gauge field interacting with only the graviton and the Higgs particle \cite{Oda4}.  

In Section 2, we present a toy model which shows spontaneous symmetry breakdown (SSB) of a global scale symmetry
\cite{Fujii, Oda3, Oda4, Oda5}. The key idea is that we begin with a scale invariant scalar-tensor gravity in the Jordan frame 
and then move to the Einstein frame. In the process of moving from the Jordan frame to the Einstein frame, 
we need to introduce a constant with mass dimension to compensate for the mass dimension of a scalar field, 
thereby triggering the SSB of the scale symmetry.  But we also point out problems of this SSB \cite{Oda5}. 
In Section 3, we briefly review a Weyl's conformal geometry. In Section 4, we present an action of a quadratic gravity 
in the Weyl geometry, for which we calculate the one-loop effective potential in the Coleman-Weinberg formalism \cite{Coleman} 
in Section 5. Section 6 is devoted to the conclusion.

\section{Spontaneous symmetry breakdown of scale symmetry} 

There is a well-known mechanism of spontaneous symmetry breakdown of a global scale symmetry
\cite{Fujii, Oda3, Oda4, Oda5}. In this section, we shall briefly review a scale invariant scalar-tensor gravity 
with two scalar fields, explain how the scale symmetry is broken spontaneously, and then point out unsatisfactory 
points of this SSB mechanism.

As a model of a scale invariant scalar-tensor gravity with two scalar fields, let us work with the following
Lagrangian density in the Jordan frame\footnote{We follow the conventions and notation adopted
in the MTW textbook \cite{MTW}.}:
\begin{eqnarray}
{\cal{L}} = \sqrt{-g} \left( \frac{1}{2} \xi \phi^2 R - \frac{1}{2} \epsilon g^{\mu\nu} \partial_\mu \phi
\partial_\nu \phi -  \frac{1}{2} g^{\mu\nu} \partial_\mu \Phi \partial_\nu \Phi 
- \frac{\lambda_1}{4} \phi^4 - \frac{\lambda_2}{2} \phi^2 \Phi^2 - \frac{\lambda_3}{4} \Phi^4 \right),
\label{J-model}
\end{eqnarray}
where $\xi$ is a constant, and $\epsilon$ takes the value $+1$ for $\phi$ being a normal field while it does $-1$
for $\phi$ being a ghost field. Moreover, $\phi$ and $\Phi$ are two distinct scalar fields, and $\lambda_i  (i = 1, 2, 3)$
are dimensionless coupling constants.  As often taken in the application for the BSM \cite{Ghilencea}, we assume that 
$\lambda_1 > 0, \lambda_3 > 0$ and $\lambda_2 < 0$, and furthermore $| \lambda_2 | \ll \lambda_1, \lambda_3 \approx
{\cal {O}} (0.1)$. The conformally invariant scalar-tensor gravity corresponds to either the case of $\xi = \frac{1}{6}$ 
and $\epsilon = - 1$ or the case of $\xi = - \frac{1}{6}$ and $\epsilon = 1$.  In this section, since we consider 
only a globally scale invariant theory, we assume $\xi > 0$ and $6 + \frac{\epsilon}{\xi} > 0$.

From this Lagrangian density, it is straightforward to derive the field equations for the metric tensor $g_{\mu\nu}$ 
and the two scalar fields $\phi, \Phi$ whose result is written as
\begin{eqnarray}
&{}& 2 \varphi G_{\mu\nu} + 2 ( g_{\mu\nu} \Box - \nabla_\mu \nabla_\nu ) \varphi = T_{\mu\nu},
\nonumber\\
&{}& \xi \phi R + \epsilon \Box \phi - \lambda_1 \phi^3 - \lambda_2 \phi \Phi^2 = 0,
\nonumber\\
&{}& \Box \Phi - \lambda_2 \phi^2 \Phi - \lambda_3 \Phi^3 = 0,
\label{Field-eq}
\end{eqnarray}
where we have defined 
\begin{eqnarray}
\varphi &=& \frac{1}{2} \xi \phi^2, \qquad G_{\mu\nu} = R_{\mu\nu} - \frac{1}{2} g_{\mu\nu} R,
\qquad
\Box \varphi = \frac{1}{\sqrt{-g}} \partial_\mu ( \sqrt{-g} g^{\mu\nu} \partial_\nu \varphi ),
\nonumber\\
T_{\mu\nu} &=& \epsilon \partial_\mu \phi \partial_\nu \phi + \partial_\mu \Phi \partial_\nu \Phi
+ g_{\mu\nu} \Biggl( - \frac{1}{2} \epsilon g^{\alpha \beta} \partial_\alpha \phi \partial_\beta \phi  
-  \frac{1}{2} g^{\alpha \beta} \partial_\alpha \Phi \partial_\beta \Phi 
\nonumber\\
&-&\frac{\lambda_1}{4} \phi^4 - \frac{\lambda_2}{2} \phi^2 \Phi^2 - \frac{\lambda_3}{4} \Phi^4 \Biggr).
\label{Various def}
\end{eqnarray}
Using these field equations, one can derive the following equation:
\begin{eqnarray}
\Box ( \varphi + \frac{\zeta^2}{2} \Phi^2 ) = 0,
\label{Derived-eq}
\end{eqnarray}
where we have defined $\zeta^{-2} \equiv 6 + \frac{\epsilon}{\xi} > 0$.

The key step for the SSB of scale invariance is to move from the Jordan frame (J-frame) to 
the Einstein frame (E-frame) by applying a conformal transformation, i.e., a local scale transformation:
\begin{eqnarray}
g_{\mu\nu} \rightarrow g^\prime_{\mu\nu}  = \Omega^2 (x) g_{\mu\nu}, \qquad
\phi \rightarrow \phi^\prime = \Omega^{-1} (x) \phi, \qquad
\Phi \rightarrow \Phi^\prime = \Omega^{-1} (x) \Phi.
\label{Conformal transf}
\end{eqnarray}
After some calculations, we can derive the transformation rule for scalar curvature \cite{Fujii}:
\begin{eqnarray}
R = \Omega^2 (x) ( R^\prime + 6 \Box^\prime f - 6 g^{\prime \mu\nu} f_\mu f_\nu ),
\label{R-transf}
\end{eqnarray}
where we have defined
\begin{eqnarray}
f = \log \Omega, \qquad f_\mu = \partial_\mu f, \qquad
\Box^\prime f = \frac{1}{\sqrt{-g^\prime}} \partial_\mu ( \sqrt{-g^\prime} g^{\prime \mu\nu}
\partial_\nu f ).
\label{Def-f}
\end{eqnarray}
Using these relations, we find that the Lagrangian density (\ref{J-model}) can be cast
to the form in a new conformal frame:
\begin{eqnarray}
{\cal{L}} &=& \sqrt{-g^\prime} \Biggl[ \frac{1}{2} \xi \phi^{\prime 2} ( R^\prime + 6 \Box^\prime f 
- 6 g^{\prime \mu\nu} f_\mu f_\nu )
- \frac{1}{2} \epsilon \Omega^{-2} g^{\prime \mu\nu} \partial_\mu ( \Omega \phi^\prime )
\partial_\nu  ( \Omega \phi^\prime )
\nonumber\\
&-&  \frac{1}{2} \Omega^{-2} g^{\prime \mu\nu} \partial_\mu ( \Omega \Phi^\prime )
\partial_\nu  ( \Omega \Phi^\prime )
- \frac{\lambda_1}{4} \phi^{\prime 4} - \frac{\lambda_2}{2} \phi^{\prime 2} \Phi^{\prime 2}
- \frac{\lambda_3}{4} \Phi^{\prime 4} \Bigr].
\label{New-frame}
\end{eqnarray}

Moving to the E-frame requires us to choose the scalar field $\phi^\prime$ to\footnote{In case of
conformal symmetry, this condition is called the ``Einstein gauge'' or ``unitary gauge''.}
\begin{eqnarray}
\phi^\prime = \frac{M_{Pl}}{\sqrt{\xi}},
\label{E-frame1}
\end{eqnarray}
where  $M_{Pl}$ is the (reduced) Planck mass defined as $M_{Pl} = \frac{1}{\sqrt{8 \pi G}} = 2.44 
\times 10^{18} GeV$ with $G$ being the Newton constant.
Then, in the E-frame, up to a total derivative, the Lagrangian density (\ref{New-frame}) reduces to the form:
\begin{eqnarray}
{\cal{L}} &=& \sqrt{-g^\prime} \Biggl( \frac{M_{Pl}^2}{2} R^\prime - \frac{1}{2} g^{\prime \mu\nu} 
\partial_\mu \sigma \partial_\nu \sigma - \frac{1}{2} g^{\prime \mu\nu} {\cal{D}}_\mu \Phi^\prime 
{\cal{D}}_\nu \Phi^\prime - \frac{\lambda_1}{4} \frac{M_{Pl}^4}{\xi^2}  
\nonumber\\
&-& \frac{\lambda_2}{2} \frac{M_{Pl}^2}{\xi} \Phi^{\prime 2} - \frac{\lambda_3}{4} \Phi^{\prime 4} \Biggr).
\label{E-model}
\end{eqnarray}
Here we have defined
\begin{eqnarray}
\Omega (x) = e^{\frac{\zeta}{M_{Pl}}  \sigma (x)}, \qquad
{\cal{D}}_\mu \Phi^\prime = \left( \partial_\mu + \frac{\zeta}{M_{Pl}} 
\partial_\mu \sigma \right) \Phi^\prime,
\label{Dilaton}
\end{eqnarray}
where a scalar field $\sigma$ is called "dilaton". 

Now, owing to our assumption $\lambda_1 > 0, \lambda_3 > 0$ and $\lambda_2 < 0$, we have a Higgs potential 
given by
\begin{eqnarray}
V(\Phi^\prime) &=& \frac{\lambda_3}{4} \Phi^{\prime 4}   
+ \frac{\lambda_2}{2} \frac{M_{Pl}^2}{\xi} \Phi^{\prime 2} + \frac{\lambda_1}{4} \frac{M_{Pl}^4}{\xi^2}
\nonumber\\
&=& \frac{\lambda_3}{4} \left( \Phi^{\prime 2} - \frac{|\lambda_2|}{\lambda_3} \frac{M_{Pl}^2}{\xi} \right)^2
+ \frac{1}{4} \left( \lambda_1 - \frac{\lambda^2_2}{\lambda_3} \right) \frac{M_{Pl}^4}{\xi^2},
\label{Higgs}
\end{eqnarray}
which determines a vacuum expectation value (VEV):
\begin{eqnarray}
\langle \Phi^\prime \rangle = \sqrt{\frac{|\lambda_2|}{\lambda_3} \frac{M_{Pl}^2}{\xi}}.
\label{VEV}
\end{eqnarray}
Expanding as $\Phi^\prime = \langle \Phi^\prime \rangle + \tilde \Phi^\prime$ with $\tilde \Phi^\prime$
being a quantum fluctuation, we have 
\begin{eqnarray}
{\cal{L}} &=& \sqrt{-g^\prime} \Biggl[ \frac{M_{Pl}^2}{2} R^\prime - \frac{1}{2} g^{\prime \mu\nu} 
\partial_\mu \sigma \partial_\nu \sigma - \frac{1}{2} g^{\prime \mu\nu} \partial_\mu \tilde \Phi^\prime 
\partial_\nu \tilde \Phi^\prime - \frac{1}{2} m^2_\Phi \tilde \Phi^{\prime 2} 
\nonumber\\
&-& \frac{\zeta}{M_{Pl}} g^{\prime \mu\nu} \tilde \Phi^\prime \partial_\mu \tilde \Phi^\prime 
\partial_\nu \sigma
- \frac{\zeta^2}{2 M_{Pl}^2} g^{\prime \mu\nu} \tilde \Phi^{\prime 2} \partial_\mu \sigma \partial_\nu \sigma
- \sqrt{\frac{\lambda_3}{2}} m_\Phi \tilde \Phi^{\prime 3} 
\nonumber\\
&-& \frac{\lambda_3}{4} \tilde \Phi^{\prime 4} 
- \frac{\lambda_1}{4} \frac{M_{Pl}^4}{\xi^2} \Biggr],
\label{E-model2}
\end{eqnarray}
where we have simplified the equations by using the relation $| \lambda_2 | \ll \lambda_1, \lambda_3 \approx
{\cal {O}} (0.1)$ and we have defined $m_\Phi = \sqrt{\frac{2 |\lambda_2|}{\xi}} M_{Pl}$. 

As is obvious from (\ref{E-model2}), the SSB of scale symmetry has occurred and as a result the scalar field 
$\tilde \Phi^\prime$ becomes massive while the ``dilaton'' $\sigma$ remains massless, which is nothing but 
a Nambu-Goldstone field. Also notice that the dilaton couples to the scalar field $\tilde \Phi^\prime$ with
derivatives which is one of characteristic features of the dilaton. To establish that $\sigma$ really plays a role
of the Nambu-Goldstone field, it is useful to derive the $\textit{dilatation current}$ associated with scale invariance, 
for which the scale factor $\Omega$ becomes a constant independent of the coordinates $x^\mu$. It is then 
convenient to consider an infinitesimal transformation given by
\begin{eqnarray}
\Omega = e^\Lambda,
\label{Infi-scale}
\end{eqnarray}
where $| \Lambda | \ll 1$. Using the Lagrangian density (\ref{J-model}) and the infinitesimal scale transformation
(\ref{Conformal transf}) with (\ref{Infi-scale}), we find that via the Noether theorem the dilatation
current $J^\mu$ reads
\begin{eqnarray}
J^\mu = \frac{1}{\zeta^2} \sqrt{-g} g^{\mu\nu} \partial_\nu \left( \varphi + \frac{\zeta^2}{2} \Phi^2 \right).
\label{dilatation}
\end{eqnarray}
The dilatation current is certainly conserved 
\begin{eqnarray}
\partial_\mu J^\mu = \frac{1}{\zeta^2} \sqrt{-g} \Box \left( \varphi + \frac{\zeta^2}{2} \Phi^2 \right)
= 0,
\label{Cons-dilatation}
\end{eqnarray}
where we have used the equation (\ref{Derived-eq}). In the E-frame, this current can be written as
\begin{eqnarray}
J^\mu = \frac{1}{2} \sqrt{-g^\prime} g^{\prime \mu\nu} \left[ \frac{2 M_{Pl}}{\zeta} \partial_\nu \sigma
+ \left( \partial_\nu + \frac{2 \zeta}{M_{Pl}} \partial_\nu \sigma \right) \Phi^{\prime 2} \right].
\label{E-dilatation}
\end{eqnarray}
Provided that one defines the dilatation charge as $Q = \int d^3 x J^0$, owing to the linear term in 
$\sigma$ its charge fails to annihilate the vacuum $| 0 \rangle$
\begin{eqnarray}
Q | 0 \rangle  \neq 0,
\label{NG}
\end{eqnarray}
which shows that the dilaton $\sigma$ is the Nambu-Goldstone boson arising from the SSB of
scale invariance.

To close this section, let us summarize the scenario of the SSB explained above and comment on its problems. 
We have started with a scale invariant gravitational theory involving two kinds of scalar fields and only
dimensionless coupling constants. In the process of moving from the J-frame to the E-frame,
we had to introduce a dimensional constant, which is the Planck mass in the present context,
to compensate for the mass dimension of the scalar field. This introduction of the Planck mass
has triggered the SSB of scale symmetry. Let us note that in the conventional scenario of the SSB, 
there is a potential inducing the SSB whereas we have no such a potential in the SSB under consideration.
Nevertheless, the very presence of a solution with dimensional constants justifies the claim 
that the present scenario of the SSB is also nothing but a spontaneous symmetry breakdown.  
Actually, this fact was explicitly verified by the dilatation charge, which does not annihilate the
vacuum due to the presence of a linear dilaton.
  
There are, however, at least two problems in this scenario of the SSB. First, it is impossible 
to apply this scenario for the conformally invariant scalar-tensor gravity, for which we must take
either $\xi = \frac{1}{6}$ and $\epsilon = - 1$ or $\xi = - \frac{1}{6}$ and $\epsilon = 1$, due to
$\zeta^{-2} \equiv 6 + \frac{\epsilon}{\xi} = 0$.  
The second problem arises from the lack of the suitable potential in the sense that
we cannot single out a solution realizing the SSB on the stability argument \cite{Fujii}.
Incidentally, though it might be possible that the cosmological argument would pick up an appropriate VEV 
of a scalar field, it is not plausible that the macroscopic physics like cosmology could determine 
a microscopic configuration such as the VEV. These two problems have been recently studied
in Ref. \cite{Oda5}.

\section{Review of Weyl conformal geometry} 

We briefly review the basic concepts and definitions of the Weyl conformal geometry.
In the Weyl geometry, the Weyl gauge transformation, which is the sum of a local scale transformation for 
a generic field $\Phi (x)$ and a gauge transformation for the Weyl gauge field $S_\mu(x)$, is defined as
\begin{eqnarray}
\Phi (x) \rightarrow \Phi^\prime (x) = e^{w \Lambda(x)} \Phi (x), \qquad
S_\mu (x) \rightarrow S^\prime_\mu (x) = S_\mu (x) - \frac{1}{f} \partial_\mu \Lambda (x),
\label{Weyl transf}
\end{eqnarray}
where $w$ is called the ``Weyl weight'', or simply ``weight'' henceforth, $f$ is the coupling constant 
for the non-compact Abelian gauge group, and $\Lambda(x)$ is a local parameter for the Weyl transformation. 
The Weyl gauge transformation for various fields is explicitly given by
\begin{eqnarray}
g_{\mu\nu} (x) &\rightarrow& g_{\mu\nu}^\prime (x) = e^{2 \Lambda(x)} g_{\mu\nu}(x), \qquad
\phi (x) \rightarrow \phi^\prime (x) = e^{- \Lambda(x)} \phi (x),  \nonumber\\
\psi (x) &\rightarrow& \psi^\prime (x) = e^{- \frac{3}{2} \Lambda(x)} \psi (x), \qquad
A_\mu (x) \rightarrow A^\prime_\mu (x) = A_\mu (x),
\label{Weyl transf 2}
\end{eqnarray}
where $g_{\mu\nu} (x)$, $\phi (x)$, $\psi (x)$ and $A_\mu (x)$ are the metric tensor, scalar, spinor,
and electromagnetic gauge fields, respectively. The covariant derivative $D_\mu$ for the Weyl gauge
transformation for a generic field $\Phi (x)$ of weight $w$ is defined as
\begin{eqnarray}
D_\mu \Phi \equiv \partial_\mu \Phi + w f S_\mu \Phi,
\label{W-cov-deriv}
\end{eqnarray}
which transforms covariantly under the Weyl transformation:
\begin{eqnarray}
D_\mu \Phi \rightarrow (D_\mu \Phi)^\prime = e^{w \Lambda(x)} D_\mu \Phi.
\label{S-cov-transf}
\end{eqnarray}

The Weyl geometry is defined as a geometry with a real symmetric metric tensor $g_{\mu\nu}
(= g_{\nu\mu})$ and a symmetric connection $\tilde \Gamma^\lambda_{\mu\nu} (= \tilde \Gamma^\lambda_{\nu\mu})$ 
which is defined as\footnote{We often use the tilde characters to express quantities belonging to the Weyl geometry.}
\begin{eqnarray}
\tilde \Gamma^\lambda_{\mu\nu} &=& \frac{1}{2} g^{\lambda\rho} \left( D_\mu g_{\nu\rho} + D_\nu g_{\mu\rho}
- D_\rho g_{\mu\nu} \right)
\nonumber\\
&=& \Gamma^\lambda_{\mu\nu} + f \left( S_\mu \delta^\lambda_\nu + S_\nu \delta^\lambda_\mu 
- S^\lambda g_{\mu\nu} \right),
\label{W-connection}
\end{eqnarray}
where 
\begin{eqnarray}
\Gamma^\lambda_{\mu\nu} \equiv \frac{1}{2} g^{\lambda\rho} \left( \partial_\mu g_{\nu\rho} 
+ \partial_\nu g_{\mu\rho} - \partial_\rho g_{\mu\nu} \right),
\label{Affine connection}
\end{eqnarray}
is the Christoffel symbol in the Riemann geometry. The most important difference between the Riemann geometry 
and the Weyl one lies in the fact that in the Riemann geometry the metric condition is satisfied 
\begin{eqnarray}
\nabla_\lambda g_{\mu\nu} \equiv \partial_\lambda g_{\mu\nu} - \Gamma^\rho_{\lambda\mu} 
g_{\rho\nu} - \Gamma^\rho_{\lambda\nu} g_{\mu\rho} = 0, 
\label{Metric cond}
\end{eqnarray}
while in the Weyl geometry we have
\begin{eqnarray}
\tilde \nabla_\lambda g_{\mu\nu} \equiv \partial_\lambda g_{\mu\nu} - \tilde \Gamma^\rho_{\lambda\mu} 
g_{\rho\nu} - \tilde \Gamma^\rho_{\lambda\nu} g_{\mu\rho}
= - 2 f S_\lambda g_{\mu\nu},
\label{W-metric cond}
\end{eqnarray}
where $\nabla_\mu$ and $\tilde \nabla_\mu$ are covariant derivatives for diffeomorphisms in the Riemann 
and Weyl geometries, respectively. Since the metric condition (\ref{Metric cond}) implies that both length 
and angle are preserved under parallel transport, Eq. (\ref{W-metric cond}) shows that only angle, 
but not length, is preserved by the Weyl connection.

The general covariant derivative for both diffeomorphisms and the Weyl gauge transformation, for instance,
for a covariant vector of weight $w$, is defined as
\begin{eqnarray}
{\cal D}_\mu V_\nu &\equiv& D_\mu V_\nu - \tilde \Gamma^\rho_{\mu\nu} V_\rho  \nonumber\\
&=& \tilde \nabla_\mu V_\nu + w f S_\mu V_\nu \nonumber\\
&=& \nabla_\mu V_\nu + w f S_\mu V_\nu - f ( S_\mu \delta^\rho _\nu + S_\nu \delta^\rho _\mu
- S^\rho g_{\mu\nu} ) V_\rho \nonumber\\
&=& \partial_\mu V_\nu + w f S_\mu V_\nu - \Gamma^\rho_{\mu\nu} V_\rho
- f ( S_\mu \delta^\rho _\nu + S_\nu \delta^\rho _\mu - S^\rho g_{\mu\nu} ) V_\rho.
\label{Gen-cov-deriv}
\end{eqnarray}
One can verify that using the general covariant derivative, the following metric condition is 
satisfied:
\begin{eqnarray}
{\cal D}_\lambda g_{\mu\nu} = 0.
\label{Gen-metric cond}
\end{eqnarray}
 Moreover, under the Weyl gauge transformation the general covariant derivative for a generic field 
 $\Phi$ of weight $w$ transforms in a covariant manner as desired:
\begin{eqnarray}
{\cal D}_\mu \Phi \rightarrow ({\cal D}_\mu \Phi)^\prime = e^{w \Lambda(x)} {\cal D}_\mu \Phi,
\label{Gen-cov-transf}
\end{eqnarray}
because the Weyl connection is invariant under the Weyl gauge transformation, i.e., 
$\tilde \Gamma^{\prime \rho}_{\mu\nu} = \tilde \Gamma^\rho_{\mu\nu}$.

As in the Riemann geometry, in the Weyl geometry one can also construct a Weyl invariant curvature
tensor $\tilde R_{\mu\nu\rho} \, ^\sigma$ via a commutator of the covariant 
derivative $\tilde \nabla_\mu$
\begin{eqnarray}
[ \tilde \nabla_\mu, \tilde \nabla_\nu ] V_\rho = \tilde R_{\mu\nu\rho} \, ^\sigma V_\sigma.
\label{Commutator}
\end{eqnarray}
Calculating this commutator, one finds that
\begin{eqnarray}
\tilde R_{\mu\nu\rho} \, ^\sigma &=& \partial_\nu \tilde \Gamma^\sigma_{\mu\rho} 
- \partial_\mu \tilde \Gamma^\sigma_{\nu\rho} + \tilde \Gamma^\alpha_{\mu\rho} \tilde \Gamma^\sigma_{\alpha\nu} 
- \tilde \Gamma^\alpha_{\nu\rho} \tilde \Gamma^\sigma_{\alpha\mu}
\nonumber\\
&=& R_{\mu\nu\rho} \, ^\sigma + 2 f \left( \delta^\sigma_{[\mu} \nabla_{\nu]} S_\rho 
- \delta^\sigma_\rho \nabla_{[\mu} S_{\nu]} - g_{\rho [\mu} \nabla_{\nu]} S^\sigma \right)
\nonumber\\
&+& 2 f^2 \left( S_{[\mu} \delta^\sigma_{\nu]} S_\rho - S_{[\mu} g_{\nu]\rho} S^\sigma
+ \delta^\sigma_{[\mu} g_{\nu]\rho} S_\alpha S^\alpha \right),
\label{W-curv-tensor}
\end{eqnarray}
where $R_{\mu\nu\rho} \, ^\sigma$ is the curvature tensor in the Riemann geometry, and
we have defined the antisymmetrization by the square bracket, e.g., $A_{[\mu} B_{\nu]} \equiv 
\frac{1}{2} ( A_\mu B_\nu - A_\nu B_\mu )$. Then, it is straightforward to prove the following identities:
\begin{eqnarray}
\tilde R_{\mu\nu\rho} \, ^\sigma = - \tilde R_{\nu\mu\rho} \, ^\sigma,  \qquad
\tilde R_{[\mu\nu\rho]} \, ^\sigma = 0, \qquad
\tilde \nabla_{[\lambda} \tilde R_{\mu\nu]\rho} \, ^\sigma = 0.
\label{W-curv-identity}
\end{eqnarray}
The curvature tensor $\tilde R_{\mu\nu\rho} \, ^\sigma$ has $26$ independent components,
twenty of which are possessed by $R_{\mu\nu\rho} \, ^\sigma$ and six by the Weyl
invariant field strength $H_{\mu\nu} \equiv \partial_\mu S_\nu - \partial_\nu S_\mu$.

From $\tilde R_{\mu\nu\rho} \, ^\sigma$ one can define a Weyl invariant 
Ricci tensor:
\begin{eqnarray}
\tilde R_{\mu\nu} &\equiv& \tilde R_{\mu\rho\nu} \, ^\rho
\nonumber\\
&=& R_{\mu\nu} + f \left( - 2 \nabla_\mu S_\nu - H_{\mu\nu} - g_{\mu\nu} \nabla_{\alpha} S^\alpha \right)
\nonumber\\
&+& 2 f^2 \left( S_\mu S_\nu - g_{\mu\nu} S_\alpha S^\alpha \right).
\label{W-Ricci-tensor}
\end{eqnarray}
Let us note that 
\begin{eqnarray}
\tilde R_{[\mu\nu]} \equiv \frac{1}{2} ( \tilde R_{\mu\nu} - \tilde R_{\nu\mu} ) = - 2 f H_{\mu\nu}.
\label{W-Ricci-tensor 2}
\end{eqnarray}
Similarly, one can define a not Weyl invariant but Weyl covariant scalar curvature:
\begin{eqnarray}
\tilde R \equiv g^{\mu\nu} \tilde R_{\mu\nu} 
= R - 6 f \nabla_\mu S^\mu - 6 f^2 S_\mu S^\mu.
\label{W-scalar-curv}
\end{eqnarray}
One finds that under the Weyl gauge transformation, $\tilde R \rightarrow \tilde R^\prime = e^{- 2 \Lambda(x)}
\tilde R$ while $\tilde \Gamma^\lambda_{\mu\nu}, \tilde R_{\mu\nu\rho} \, ^\sigma$ and $\tilde R_{\mu\nu}$
are all invariant.

Even in the Weyl geometry, it is possible to write out a generalization of the Gauss-Bonnet topological 
invariant which can be described as
\begin{eqnarray}
I_{GB} &\equiv& \int d^4 x \sqrt{-g} \, \epsilon^{\mu\nu\rho\sigma} \epsilon_{\alpha\beta\gamma\delta} \, 
\tilde R_{\mu\nu} \, ^{\alpha\beta} \tilde R_{\rho\sigma} \, ^{\gamma\delta}    
\nonumber\\
&=& - 2 \int d^4 x \sqrt{-g} \,  \left( \tilde R_{\mu\nu\rho\sigma} \tilde R^{\rho\sigma\mu\nu} 
- 4 \tilde R_{\mu\nu} \tilde R^{\nu\mu} + \tilde R^2 - 12 f^2 H_{\mu\nu} H^{\mu\nu} \right)
\nonumber\\
&=& - 2 \int d^4 x \sqrt{-g} \,  \left( R_{\mu\nu\rho\sigma} R^{\mu\nu\rho\sigma} 
- 4 R_{\mu\nu} R^{\mu\nu} + R^2 \right).   
\label{GB}
\end{eqnarray}

We close this section by discussing a spinor field as an example of matter fields in the Weyl
geometry \cite{Shirafuji, Hayashi}.  As is well known, to describe a spinor field it is necessary to introduce 
the vierbein $e^a _\mu$, which is defined as
\begin{eqnarray}
g_{\mu\nu} = \eta_{ab} e^a _\mu e^b _\nu,
\label{Vierbein}
\end{eqnarray}
where $a, b, \cdots$ are local Lorentz indices taking $0, 1, 2, 3$ and $\eta_{ab} = diag ( - 1, 1, 1, 1)$.

Now the metric condition (\ref{Gen-metric cond}) takes the form 
\begin{eqnarray}
{\cal D}_\mu e^a _\nu \equiv D_\mu e^a _\nu + \tilde \omega^a \, _{b \mu} e^b _\nu 
- \tilde \Gamma^\rho_{\mu\nu} e^a _\rho = 0,
\label{Gen-vierbein cond}
\end{eqnarray}
where the general covariant derivative is extended to include the local Lorentz transformation whose
gauge connection is the spin connection $\tilde \omega^a \, _{b \mu}$ of weight $0$ in the Weyl
geometry, and $D_\mu e^a _\nu = \partial_\mu e^a _\nu + f S_\mu e^a _\nu$ since the vierbein 
$e^a _\mu$ has weight $1$. Solving the metric condition (\ref{Gen-vierbein cond}) leads to the
expression of the spin connection in the Weyl geometry
\begin{eqnarray}
\tilde \omega_{a b \mu} = \omega_{a b \mu} + f e^c _\mu ( \eta_{ac} S_b - \eta_{bc} S_a ),
\label{spin connection}
\end{eqnarray}
where $\omega_{a b \mu}$ is the spin connection in the Riemann geometry and we have defined 
$S_a \equiv e^\mu _a S_\mu$. Then, the general covariant 
derivative for a spinor field $\Psi$ of weight $- \frac{3}{2}$ reads
\begin{eqnarray}
{\cal D}_\mu \Psi = D_\mu \Psi + \frac{i}{2} \tilde \omega_{a b \mu} S^{a b}  \Psi,
\label{spinor CD}
\end{eqnarray}
where $D_\mu \Psi = \partial_\mu \Psi - \frac{3}{2} f S_\mu \Psi$ and the Lorentz generator $S^{a b}$
for a spinor field is defined as $S^{a b} = \frac{i}{4} [ \gamma^a, \gamma^b ]$. Here we define the gamma
matrices to satisfy the Clifford algebra $\{ \gamma^a, \gamma^b \} = - 2 \eta^{ab}$.
Since the spin connection $\tilde \omega^a \, _{b \mu}$ has weight $0$, the covariant
derivative ${\cal D}_\mu \Psi$  transforms covariantly under the Weyl gauge transformation
\begin{eqnarray}
{\cal D}_\mu \Psi \rightarrow ( {\cal D}_\mu \Psi )^\prime = e^{- \frac{3}{2} \Lambda(x)} 
{\cal D}_\mu \Psi.
\label{spinor covariance}
\end{eqnarray}

Then, the Lagrangian density for a massless Dirac spinor field is of form
\begin{eqnarray}
{\cal L} = \frac{i}{2} e \ e^\mu _a ( \bar \Psi \gamma^a {\cal D}_\mu \Psi 
- {\cal D}_\mu \bar \Psi \gamma^a \Psi ),
\label{spinor Lag}
\end{eqnarray}
where $e \equiv \sqrt{-g}, \bar \Psi \equiv \Psi^\dagger \gamma^0$, and ${\cal D}_\mu \bar \Psi$
is given by
\begin{eqnarray}
{\cal D}_\mu \bar \Psi = D_\mu \bar \Psi - \bar \Psi \frac{i}{2} \tilde \omega_{a b \mu} S^{a b}.
\label{spinor CD2}
\end{eqnarray}
Inserting Eqs. (\ref{spinor CD}) and (\ref{spinor CD2}) to the Lagrangian density (\ref{spinor Lag}), 
we find that  
\begin{eqnarray}
{\cal L} &=& \frac{i}{2} e \Bigl[ e^\mu _a  \left( \bar \Psi \gamma^a \partial_\mu \Psi 
- \partial_\mu \bar \Psi \gamma^a \Psi + \frac{i}{2} \omega_{b c \mu} \bar \Psi \{ \gamma^a,
S^{bc} \} \Psi \right)
\nonumber\\
&+& \frac{i}{2} f ( \eta_{ab} S_c - \eta_{ac} S_b ) \bar \Psi \{ \gamma^a, S^{bc} \} \Psi \Bigr].
\label{spinor Lag2}
\end{eqnarray}
The last term identically vanishes owing to the relation 
\begin{eqnarray}
\{ \gamma^a, S^{bc} \} = - \varepsilon^{abcd} \gamma_5 \gamma_d,
\label{gamma rel}
\end{eqnarray}
where we have defined as $\gamma_5 = i \gamma^0 \gamma^1 \gamma^2 \gamma^3$ and 
$\varepsilon^{0123} = +1$. Thus, as is well known, the Weyl gauge field $S_\mu$ does not couple minimally 
to a spinor field $\Psi$.  Technically speaking, it is the absence of imaginary unit $i$ in the 
covariant derivative $D_\mu \Psi = \partial_\mu \Psi - \frac{3}{2} f S_\mu \Psi$ that induced this
decoupling of the Weyl gauge field from the spinor field. Without the imaginary unit, the terms including 
the Weyl gauge field cancel out each other in Eq. (\ref{spinor Lag}). In a similar manner, we can prove that 
the Weyl gauge field does not couple to a gauge field either such as the electromagnetic potential $A_\mu$. 
On the other hand, the Weyl gauge field can couple to a scalar field such as the Higgs field as well as a graviton. 
In such a situation, we cannot help identifying the Weyl gauge field with an elementary particle that constitutes 
dark matter. It seems that the Weyl gauge theory was rejected as a unified theory of gravitation and electromagnetism
but it has revived as a geometrical theory which predicts the existence of dark matter.

\section{Quadratic gravity in Weyl geometry} 

In this section, we will present a gravitational theory on the basis of the Weyl geometry outlined 
in the previous section. It is of interest to notice that if only the metric tensor is allowed to use for 
the construction of a gravitational action, the action invariant under the Weyl transformation must be 
of form of quadratic gravity, but not be of the Einstein-Hilbert type. Using the topological invariant (\ref{GB}), 
one can write out a general action of quadratic gravity, which is invariant under the Weyl transformation, as follows:
\begin{eqnarray}
S_{QG} = \int d^4 x \sqrt{-g} \left[ - \frac{1}{2 \xi^2}  \tilde C_{\mu\nu\rho\sigma} \tilde C^{\mu\nu\rho\sigma} 
+ \frac{\lambda}{4 !} \tilde R^2 \right] \equiv \int d^4 x \sqrt{-g} \, {\cal{L}}_{QG},
\label{QG}
\end{eqnarray}
where $\xi$ and $\lambda$ are dimensionless coupling constants. And a generalization of the conformal tensor, 
$\tilde C_{\mu\nu\rho\sigma}$, in the Weyl geometry is defined as in $C_{\mu\nu\rho\sigma}$ in the Riemann geometry:
\begin{eqnarray}
\tilde C_{\mu\nu\rho\sigma} &\equiv& \tilde R_{\mu\nu\rho\sigma} - \frac{1}{2} \left( g_{\mu\rho} \tilde R_{\nu\sigma}
+ g_{\nu\sigma} \tilde R_{\mu\rho} -  g_{\mu\sigma} \tilde R_{\nu\rho} - g_{\nu\rho} \tilde R_{\mu\sigma} \right)
+ \frac{1}{6} \left( g_{\mu\rho} g_{\nu\sigma} - g_{\mu\sigma} g_{\nu\rho} \right) \tilde R
\nonumber\\
&=& C_{\mu\nu\rho\sigma} + f \left[ - g_{\rho\sigma} H_{\mu\nu} + \frac{1}{2} \left( g_{\mu\rho} H_{\nu\sigma}
+  g_{\nu\sigma} H_{\mu\rho} - g_{\mu\sigma} H_{\nu\rho} - g_{\nu\rho} H_{\mu\sigma} \right) \right].
\label{Conformal tensor}
\end{eqnarray}
This conformal tensor in the Weyl geometry has the following properties:
\begin{eqnarray}
\tilde C_{\mu\nu\rho\sigma} = - \tilde C_{\nu\mu\rho\sigma}, \qquad 
\tilde C_{\mu\nu\rho} \, ^\nu = 0, \qquad 
\tilde C_{\mu\nu\rho} \, ^\rho = - 4 f H_{\mu\nu}.
\label{Conformal tensor 2}
\end{eqnarray}

Next, by introducing a scalar field $\phi$ and using the classical equivalence, let us rewrite $\tilde R^2$ in the action 
(\ref{QG}) in the form of the scalar-tensor gravity plus $\lambda \phi^4$ interaction \cite{Ghilencea1, Ghilencea2} 
whose Lagrangian density takes the form
\begin{eqnarray}
\frac{1}{\sqrt{-g}} {\cal{L}}_{QG} &=& - \frac{1}{2 \xi^2}  \tilde C_{\mu\nu\rho\sigma} \tilde C^{\mu\nu\rho\sigma} 
+ \frac{\lambda}{12} \phi^2 \tilde R - \frac{\lambda}{4 !} \phi^4 
\nonumber\\
&=& - \frac{1}{2 \xi^2}  \tilde C_{\mu\nu\rho\sigma} \tilde C^{\mu\nu\rho\sigma} 
+ \frac{1}{12} \phi^2 \tilde R - \frac{\lambda_\phi}{4 !} \phi^4 
\nonumber\\
&=& - \frac{1}{2 \xi^2}  C_{\mu\nu\rho\sigma} C^{\mu\nu\rho\sigma} 
+ \frac{1}{12} \phi^2 R - \frac{\lambda_\phi}{4 !} \phi^4 - \frac{3 f^2}{\xi^2} H_{\mu\nu}^2 
\nonumber\\
&-& \frac{1}{2} \phi^2 ( f \nabla_\mu S^\mu + f^2 S_\mu S^\mu ),  
\label{QG 2}
\end{eqnarray}
where in the second equality we have redefined $\sqrt{\lambda} \phi \rightarrow \phi$ and set 
$\lambda = \frac{1}{\lambda_\phi}$. It is straightforward to write down a standard model (SM) or physics
beyond the standard model (BSM) action which is invariant under the Weyl transformation, but we will omit to 
do it in this article and present the detail in a separate publication.

\section{Emergence of Planck scale}

At low energies, general relativity (GR) describes various gravitational and astrophysical phenomena neatly,
so the Weyl invariant Lagrangian density (\ref{QG 2}) of quadratic gravity should be reduced to that of GR at low
energies. To do that, we need to break the Weyl symmetry at any rate by some method. One method is to 
appeal to the procedure of spontaneous symmetry breakdown (SSB) explained in terms of a toy model in Section 2.
However, as emphasized there, since there is no potential to induce this SSB in the theory, we have no idea which
solution we should pick up among many of configurations from the stability argument. 

The other simple procedure is to take a gauge condition for the Weyl transformation such that $\phi = \phi_0$ where $\phi_0$ 
is a certain constant \cite{Hayashi, Smolin, Cesare, Ghilencea1, Ghilencea2}. However, $\phi_0$ is a free parameter 
which is not fixed from the stability argument of the potential either so it is not clear why we choose a specific value 
$\phi_0 \sim M_{Pl}$.     

In this article, we would like to look for an alternative possibility by considering a conformally invariant gravitational theory 
where the scalar field $\phi$ acquires a vacuum expectation value (VEV) as a result of instabilities in the full quantum theory 
including quantum corrections from gravity. It is natural to conjecture that quantum gravity plays a role in
generating the Planck mass scale dynamically since effects of quantum gravity are more dominant than the other interactions
around the Planck mass. Technically speaking, what we expect is that after quantum corrections of gravitational fields 
are taken into consideration the effective potential has a form favoring the specific VEV, $\phi_0 \sim M_{Pl}$ 
\cite{Oda1, Oda2, Oda3}.

To this aim, let us first expand the scalar field and the metric around a classical field $\phi_c$ and a flat Minkowski
metric $\eta_{\mu\nu}$ like \cite{Oda1, Oda2, Oda3}
\begin{eqnarray}
\phi = \phi_c + \varphi,    \quad 
g_{\mu\nu} = \eta_{\mu\nu} + \xi h_{\mu\nu},
\label{Expansion}
\end{eqnarray}
where we take $\phi_c$ to be a constant since we are interested in the effective potential depending on
the constant $\phi_c$. Next, since we wish to calculate the one-loop effective potential, we will
derive only quadratic terms in quantum fields from the classical Lagrangian density (\ref{QG 2}).
Then, the Lagrangian density corresponding to the conformal tensor squared takes the form
\begin{eqnarray}
{\cal L}_C \equiv - \frac{1}{2 \xi^2}  \sqrt{-g} \, C_{\mu\nu\rho\sigma} C^{\mu\nu\rho\sigma} 
= - \frac{1}{4} h^{\mu\nu} P^{(2)}_{\mu\nu, \rho\sigma} \Box^2 h^{\rho\sigma},
\label{Weyl Lagr}
\end{eqnarray}
where $P^{(2)}_{\mu\nu, \rho\sigma}$ is the projection operator for spin-2 modes\footnote{We follow
the definition of projection operators in \cite{Nakasone1, Nakasone2}.} and $\Box \equiv \eta^{\mu\nu}
\partial_\mu \partial_\nu$. In a similar manner, the Lagrangian density corresponding to the scalar-tensor
gravity in Eq. (\ref{QG 2}) reads
\begin{eqnarray}
{\cal L}_{ST} &\equiv& \sqrt{-g} \, \frac{1}{12} \phi^2 R
\nonumber\\
&=& \frac{1}{48} \xi^2 \phi^2_c h^{\mu\nu} \left( P^{(2)}_{\mu\nu, \rho\sigma} 
- 2 P^{(0, s)}_{\mu\nu, \rho\sigma} \right) \Box h^{\rho\sigma} 
- \frac{1}{6} \xi \phi_c \varphi \left( \eta_{\mu\nu} - \frac{1}{\Box} \partial_\mu \partial_\nu \right) 
\Box h^{\mu\nu}.
\label{ST-Lagr}
\end{eqnarray}
The remaining Lagrangian density can be evaluated in a similar way and consequently all the quadratic terms 
in (\ref{QG 2}) are summarized to
\begin{eqnarray}
{\cal L}_{QG} &=& \frac{1}{4} h^{\mu\nu} \left[ \left( - \Box + \frac{1}{12} \xi^2 \phi^2_c \right)
P^{(2)}_{\mu\nu, \rho\sigma} - \frac{1}{6} \xi^2 \phi^2_c P^{(0, s)}_{\mu\nu, \rho\sigma} \right]
\Box h^{\rho\sigma} 
\nonumber\\
&-& \frac{1}{6} \xi \phi_c \varphi \left( \eta_{\mu\nu} 
- \frac{1}{\Box} \partial_\mu \partial_\nu \right) \Box h^{\mu\nu}
- \frac{\lambda_\phi}{4} \phi_c^2 \varphi^2 -  \frac{\lambda_\phi}{12} \xi \phi_c^3 h \varphi
+ \frac{1}{96} \lambda_\phi \xi^2 \phi_c^4 h_{\mu\nu}^2 
\nonumber\\
&-& \frac{1}{192} \lambda_\phi \xi^2 \phi_c^4 h^2 - \frac{1}{4} H^{\prime \ 2}_{\mu\nu} 
- \frac{1}{24} \xi^2 \phi_c^2 S^\prime_\mu S^{\prime \mu} - \frac{1}{2} \varphi \Box \varphi,
\label{Total-Lagr}
\end{eqnarray}
where we have defined $h= \eta^{\mu\nu} h_{\mu\nu}$ and set $S^\prime_\mu = \frac{2 \sqrt{3} f}{\xi}
 ( S_\mu - \frac{1}{f \phi_c} \partial_\mu \varphi)$  and $H^\prime_{\mu\nu} = \partial_\mu S^\prime_\nu 
- \partial_\nu S^\prime_\mu$. In what follows, we will assume that
\begin{eqnarray}
\lambda_\phi \propto \xi^4 \ll 1,
\label{Assumption}
\end{eqnarray}
and drop all the terms involving $\lambda_\phi$. We will prove later that our assumption (\ref{Assumption})
is self-consistent and there are no large logarithms.

At this point, it is convenient to use the York decomposition for the metric fluctuation field $h_{\mu\nu}$
\cite{York}:
\begin{eqnarray}
h_{\mu\nu} &=& h_{\mu\nu}^{TT} + \partial_\mu \xi_\nu + \partial_\nu \xi_\mu + \partial_\nu \partial_\nu \sigma
- \frac{1}{4} \eta_{\mu\nu} \Box \sigma + \frac{1}{4} \eta_{\mu\nu} h \nonumber\\
&=& h_{\mu\nu}^{TT} + \partial_\mu \xi_\nu + \partial_\nu \xi_\mu + \partial_\nu \partial_\nu \sigma
+ \frac{1}{4} \theta_{\mu\nu} s + \frac{1}{4} \omega_{\mu\nu} w,
\label{York}
\end{eqnarray}
where $h_{\mu\nu}^{TT}$ is both transverse and traceless, and $\xi_\mu$ is transverse:
\begin{eqnarray}
\partial^\mu h_{\mu\nu}^{TT} = \eta^{\mu\nu} h_{\mu\nu}^{TT} = \partial^\mu \xi = 0.
\label{TT}
\end{eqnarray}
Moreover, we have defined 
\begin{eqnarray}
s = h - \Box \sigma, \quad  w = h + 3 \Box \sigma, \quad 
\theta_{\mu\nu} =  \eta_{\mu\nu} - \frac{1}{\Box} \partial_\mu \partial_\nu, \quad
\omega_{\mu\nu} = \frac{1}{\Box} \partial_\mu \partial_\nu.
\label{s&w}
\end{eqnarray}
One advantage of the York decomposition (\ref{York}) is that each term corresponds to 
the degree of freedom with the definite spin as seen in the following relations:
\begin{eqnarray}
P_{\mu\nu}^{(2) \rho\sigma} h_{\rho\sigma} &=& h_{\mu\nu}^{TT}, \qquad 
P_{\mu\nu}^{(1) \rho\sigma} h_{\rho\sigma} = \partial_\mu \xi_\nu + \partial_\nu \xi_\mu,
\nonumber\\
P_{\mu\nu}^{(0, s) \rho\sigma} h_{\rho\sigma} &=&  \frac{1}{4} \theta_{\mu\nu} s, \qquad
P_{\mu\nu}^{(0, w) \rho\sigma} h_{\rho\sigma} =  \frac{1}{4} \omega_{\mu\nu} w.
\label{spin-proj}
\end{eqnarray}
Using these relations and our assumption (\ref{Assumption}), the Lagrangian density (\ref{Total-Lagr}) reads
\begin{eqnarray}
{\cal L}_{QG} = \frac{1}{4} h^{TT \mu\nu} ( - \Box + m^2 ) \Box h^{TT}_{\mu\nu}
- \frac{1}{2} \varphi^\prime \Box \varphi^\prime
- \frac{1}{4} H^{\prime \ 2}_{\mu\nu} - \frac{m^2}{2} S^\prime_\mu S^{\prime \mu},
\label{Total-Lagr2}
\end{eqnarray}
where we have put $m^2 = \frac{1}{12} \xi^2 \phi_c^2$ and $\varphi^\prime = \varphi
- \frac{\sqrt{3} m}{4} s$.
 
Now let us calculate the functional Jacobian associated with the change of variables, $h_{\mu\nu}
\rightarrow (h^{TT}_{\mu\nu}, \xi_\mu, s, w)$. To do that, we will use the relation \cite{Percacci}
\begin{eqnarray}
1 = \int {\cal{D}} h_{\mu\nu} e^{ - {\cal{G}}(h, h) },
\label{Measure}
\end{eqnarray}
where ${\cal{G}}(h, h)$ is an inner product in the space of symmetric rank-2 tensors:
\begin{eqnarray}
{\cal{G}}(h, h) &=& \int d^4 x ( h_{\mu\nu} h^{\mu\nu} + \frac{a}{2} h^2 )  \nonumber\\
&=& \int d^4 x \left[ (h^{TT}_{\mu\nu})^2 - 2 \xi_\mu \Box \xi^\mu + \frac{3 (3a + 2)}{32} 
s^{\prime 2} + \frac{2 a + 1}{8 (3a + 2)} w^2 \right],
\label{G}
\end{eqnarray}
where $a$ is an arbitrary constant and we have defined $s^\prime = s + \frac{3}{3a + 2} w$. Thus, 
the functional Jacobian $J$ which is defined as
\begin{eqnarray}
{\cal{D}} h_{\mu\nu} = J  {\cal{D}} h^{TT}_{\mu\nu} {\cal{D}} \xi_\mu {\cal{D}} s^\prime {\cal{D}} w,
\label{J}
\end{eqnarray}
is given by
\begin{eqnarray}
J  = ( \det{}_\xi \Box )^{\frac{1}{2}}.
\label{J2}
\end{eqnarray}

Next let us set up the gauge-fixing conditions. For diffeomorphisms and the Weyl transformation, we adopt 
gauge conditions, respectively
\begin{eqnarray}
\partial^\nu h_{\mu\nu} = \Box \xi_\mu + \frac{1}{4} \partial_\mu w = 0, \qquad
\partial_\mu S^{\prime \mu} = 0.
\label{Gauge-diffo}
\end{eqnarray}
The corresponding FP ghost terms are respectively calculated to
\begin{eqnarray}
\det \Delta_{FP}^{(GCT)} = \det ( \Box \delta_\mu^\nu + \partial_\mu \partial^\nu ), \qquad
\det \Delta_{FP}^{(Weyl)} = \det ( \Box ).
\label{FP}
\end{eqnarray}
Then, the partition function of the present theory is given by
\begin{eqnarray}
Z [ \phi_c ] &=& \int {\cal{D}} g_{\mu\nu} {\cal{D}} \phi {\cal{D}} S_\mu \det \Delta_{FP}^{(GCT)}
\det \Delta_{FP}^{(Weyl)} \delta(\partial^\nu h_{\mu\nu}) \delta(\partial_\mu S^{\prime \mu}) 
\nonumber\\
&\times& exp \ i \int d^4 x \left[ \frac{1}{4} h^{TT \mu\nu} ( - \Box + m^2 ) \Box h^{TT}_{\mu\nu}
- \frac{1}{2} \varphi^\prime \Box \varphi^\prime
- \frac{1}{4} H^{\prime \ 2}_{\mu\nu} - \frac{m^2}{2} S^\prime_\mu S^{\prime \mu} \right]
\nonumber\\
&=& \int {\cal{D}} h^{TT}_{\mu\nu} {\cal{D}} \xi_\mu {\cal{D}} s^\prime {\cal{D}} w
{\cal{D}} \varphi^\prime {\cal{D}} S^\prime_\mu ( \det{}_\xi \Box )^{\frac{1}{2}}
\det ( \Box \delta_\mu^\nu + \partial_\mu \partial^\nu )  \det ( \Box )
\nonumber\\
&\times& \delta(\Box \xi_\mu + \frac{1}{4} \partial_\mu w) \delta(\partial_\mu S^{\prime \mu}) 
exp \ i \int d^4 x \Biggl[ \frac{1}{4} h^{TT \mu\nu} ( - \Box + m^2 ) \Box h^{TT}_{\mu\nu}
\nonumber\\
&-& \frac{1}{2} \varphi^\prime \Box \varphi^\prime
- \frac{1}{2} S^{\prime \mu} ( - \Box + m^2 ) S^\prime_\mu + \frac{1}{2} \partial_\mu S^{\prime \mu} )^2 
\Biggr]
\nonumber\\
&=& \frac{\det ( \Box \delta_\mu^\nu + \partial_\mu \partial^\nu )  \det ( \Box )}
{( \det{}_\xi \Box )^{\frac{1}{2}} ( \det{}_{\varphi^\prime} \Box )^{\frac{1}{2}}
( \det{}_{h^{TT}} ( - \Box + m^2 ) \Box )^{\frac{1}{2}}
( \det{}_{S^\prime} ( - \Box + m^2 ) )^{\frac{1}{2}}}.
\label{Partition}
\end{eqnarray}

Using the partition function (\ref{Partition}), we can evaluate the one-loop effective action
by integrating out quantum fluctuations.  Then, up to a classical potential, recalling the definition 
$m^2 = \frac{1}{12} \xi^2 \phi_c^2$, the effective action $\Gamma [\phi_c]$ reads
\begin{eqnarray}
\Gamma [\phi_c] = - i \log Z [\phi_c] = i \frac{5 + 3}{2} \log \mathrm{det} 
\left( - \Box + \frac{1}{12} \xi^2 \phi^2_c \right).
\label{EA}
\end{eqnarray}
Here some remarks are in order. First, in this expression, the factors $5$ and $3$ come from the fact that 
a massive spin-2 state and a massive spin-1 Weyl gauge field possess five and three physical degrees of freedom, 
respectively. Second, let us note that we have ignored the part of the effective action which is independent 
of $\phi_c$ since it never gives us the effective potential for $\phi_c$. 

To calculate $\Gamma [\phi_c]$, we will proceed step by step: First, let us note that $\Gamma [\phi_c]$ can be 
rewritten as follows:
\begin{eqnarray}
\Gamma [\phi_c] &=& 4 i \, \mathrm{Tr} \log \left( - \Box + \frac{1}{12} \xi^2 \phi^2_c \right)
\nonumber\\
&=& 4 i \int d^4 x \, \langle x|  \log \left( - \Box + \frac{1}{12} \xi^2 \phi^2_c \right) | x \rangle
\nonumber\\
&=& 4 i \int d^4 x \int \frac{d^4 k}{(2 \pi)^4} \, \langle x|  \log \left( - \Box + \frac{1}{12} \xi^2 \phi^2_c \right) 
| k \rangle \langle k | x \rangle
\nonumber\\
&=& 4 i (VT) \int \frac{d^4 k}{(2 \pi)^4} \log \left( k^2 + \frac{1}{12} \xi^2 \phi^2_c \right)
\nonumber\\
&=& 4 (VT) \frac{\Gamma(- \frac{d}{2})}{(4 \pi)^{\frac{d}{2}}} \left( \frac{1}{12} \xi^2 \phi^2_c \right)^{\frac{d}{2}},
\label{EA2}
\end{eqnarray}
where $(VT)$ denotes the space-time volume and in the last equality we have used the Wick rotation and 
the dimensional regularization.

Next, let us evaluate the $\Gamma [\phi_c]$ in terms of the modified minimal subtraction scheme. In this scheme, 
the $\frac{1}{\varepsilon}$ poles (where $\varepsilon \equiv 4 - d$) together with the Euler-Mascheroni constant
$\gamma$ and $\log (4 \pi)$ are subtracted and then replaced with $\log M^2$ where $M$ is an arbitrary
mass parameter which is introduced to make the final equation dimensionally correct \cite{Peskin}. 
By subtracting the $\frac{1}{\varepsilon}$ pole, (\ref{EA2}) is reduced to the form
\begin{eqnarray}
- \frac{1}{VT} \Gamma [\phi_c] &=& - 4 \frac{\Gamma( 2 - \frac{d}{2} )}{\frac{d}{2} (\frac{d}{2} - 1)}
\frac{1}{(4 \pi)^{\frac{d}{2}}} \left( \frac{1}{12} \xi^2 \phi^2_c \right)^{\frac{d}{2}}
\nonumber\\
&=& - \frac{4}{2 (4 \pi)^2} \left( \frac{1}{12} \xi^2 \phi^2_c \right)^2 \left[ \frac{2}{\varepsilon} - \gamma
+ \log (4 \pi)  - \log \left(\frac{1}{12} \xi^2 \phi^2_c \right) + \frac{3}{2} \right]
\nonumber\\
&\rightarrow& \frac{2}{(4 \pi)^2} \left( \frac{1}{12} \xi^2 \phi^2_c \right)^2 
\left[ \log \left(\frac{\xi^2 \phi^2_c}{12 M^2} \right) - \frac{3}{2} \right].
\label{EA3}
\end{eqnarray}
Then, the one-loop effective potential will be of form\footnote{At first sight, the existence of the
$c_2 \phi^2$ might appear to be strange, but this term in fact emerges in the cutoff regularization. Note that
the only logarithmically divergent term, but not quadratic divergent one, arises in the dimensional regularization.}
\begin{eqnarray}
V_{eff}^{(1)} (\phi_c) = c_1 + c_2 \phi^2 + \frac{1}{1152 \pi^2} \xi^4 \phi^4_c \log \left(\frac{\phi^2_c}{c_3}\right),
\label{EP}
\end{eqnarray}
where $c_i (i = 1, 2, 3)$ are constants to be determined by the renormalization conditions:
\begin{eqnarray}
\left. V_{eff}^{(1)} \right\vert_{\phi_c = 0} = \left. \frac{d^2 V_{eff}^{(1)}}{d \phi^2_c} \right\vert_{\phi_c = 0} 
= \left. \frac{d^4 V_{eff}^{(1)}}{d \phi^4_c} \right\vert_{\phi_c = \mu} = 0,
\label{Ren-cond}
\end{eqnarray}
where $\mu$ is the renormalization mass. As a result, we have the one-loop effective potential
\begin{eqnarray}
V_{eff}^{(1)} (\phi_c) = \frac{1}{1152 \pi^2} \xi^4 \phi^4_c \left( \log \frac{\phi^2_c}{\mu^2} - \frac{25}{6} \right).
\label{EP2}
\end{eqnarray}
Finally, by adding the classical potential we can arrive at the effective potential in the one-loop approximation
\begin{eqnarray}
V_{eff} (\phi_c) = \frac{\lambda_\phi}{4 !} \phi^4_c + \frac{1}{1152 \pi^2} \xi^4 \phi^4_c 
\left( \log \frac{\phi^2_c}{\mu^2} - \frac{25}{6} \right).
\label{EP3}
\end{eqnarray}

It is easy to see that this effective potential has a minimum at $\phi_c = \langle \phi \rangle$ away from the origin 
where the effective potential, $V_{eff} (\langle \phi \rangle)$, is negative. Since the renormalization mass $\mu$ 
is arbitrary, we will choose it to be the actual location of the minimum, $\mu = \langle \phi \rangle$ \cite{Coleman}:
\begin{eqnarray}
V_{eff} (\phi_c) = \frac{\lambda_\phi}{4 !} \phi^4_c + \frac{1}{1152 \pi^2} \xi^4 \phi^4_c 
\left( \log \frac{\phi^2_c}{\langle \phi \rangle^2} - \frac{25}{6} \right).
\label{EP4}
\end{eqnarray}
Since $\phi_c = \langle \phi \rangle$ is defined to be the minimum of $V_{eff}$, we deduce
\begin{eqnarray}
0 &=& \left. \frac{d V_{eff}}{d \phi_c} \right\vert_{\phi_c = \langle \phi \rangle}
\nonumber\\ 
&=& \left( \frac{\lambda_\phi}{6} - \frac{11}{864 \pi^2} \xi^4 \right) \langle \phi \rangle^3,
\label{Min-cond}
\end{eqnarray}
or equivalently, 
\begin{eqnarray}
\lambda_\phi = \frac{11}{144 \pi^2} \xi^4.
\label{Min-cond2}
\end{eqnarray}
This relation is similar to $\lambda = \frac{33}{8 \pi^2} e^4$ in case of the scalar QED in 
Ref. \cite{Coleman}, so as in that paper, the perturbation theory holds for very small $\xi$ as well.

The substitution of Eq. (\ref{Min-cond2}) into $V_{eff}$ in (\ref{EP4}) leads to
\begin{eqnarray}
V_{eff} (\phi_c) = \frac{1}{1152 \pi^2} \xi^4 \phi^4_c 
\left( \log \frac{\phi^2_c}{\langle \phi \rangle^2} - \frac{1}{2} \right).
\label{EP5}
\end{eqnarray}
Thus, the effective potential is now parametrized in terms of $\xi$ and $\langle \phi \rangle$
instead of $\xi$ and $\lambda_\phi$; it is nothing but the well-known "dimensional transmutation"
, i.e., a dimensionless coupling constant $\lambda_\phi$ is traded for a dimensional quantity 
$\langle \phi \rangle$ via symmetry breakdown of the $\it{local}$ Weyl symmetry.   

Hence, from the classical Lagrangian density (\ref{QG 2}) of quadratic gravity, via dimensional transmutation,
the Einstein-Hilbert term for GR is induced in such a way that the Planck mass $M_{Pl}$ is given by
\begin{eqnarray}
M_{Pl}^2 = \frac{1}{6} \langle \phi \rangle^2.
\label{Planck mass}
\end{eqnarray}
At the same time, the Weyl gauge field becomes massive by 'eating' the scalar graviton $s$ and a part of
the dilaton $\varphi$ whose magnitude of mass is given 
\begin{eqnarray}
m_S^2 = \frac{1}{12} \xi^2 \langle \phi \rangle^2 = \frac{1}{2} \xi^2 M_{Pl}^2.
\label{Gauge mass}
\end{eqnarray}
As long as the perturbation theory is concerned, the coupling constant $\xi$ must take a small value,
$\xi \ll 1$. At the low energy region satisfying $E \ll m_S$, we can integrate over the massive Weyl gauge field,
and consequently not only we would have GR with the SM but also the second clock effect has no physical
effects at low energies.

\section{Conclusions}

Shortly after Einstein constructed general relativity (GR) in 1915, Weyl has advocated a generalization in that the very 
notion of length becomes path-dependent. In Weyl's theory, even if the lightcones retain the fundamental role as in GR, 
there is no absolute meaning of scales for space-time, so the metric is defined only up to proportionality. 
It is this property that we have a scale symmetry prohibiting the appearance of any dimensionful parameters and coupling constants 
in the Weyl theory. The main complaint against the Weyl's idea is that it inevitably leads to the so-called "second clock 
effect": The rate where any clock measures would depend on its history. Since the second clock effect has not been 
observed by experiments, the Weyl theory might make no sense as a classical theory.\footnote{The second clock problem
and its resolution have been recently discussed in Ref. \cite{Oda4}.}  

However, viewed as a quantum field theory, the Weyl theory is a physically consistent theory and provides us with
a natural playground for constructing conformally invariant quantum field theories as shown in this article.\footnote{We
have already contructed the other scale invariant gravitational models \cite{Oda6, Oda7, Oda8}.}
Requiring the invariance under Weyl transformation is so strong that only quadratic curvature terms are allowed to
exist in a classical action, which should be contrasted with the situation of GR where any number of curvature terms could be 
in principle the candidate of a classical action only ifs we require the action to be invariant under diffeomorphisms.

Of course, we have a serious problem to be solved in future; the problem of unitarity. The lack of perturbative
unitarity is a common problem in the higher derivative gravity like the Weyl theory \cite{Stelle, Tonin, Fradkin}. 
However, it is expected that the Weyl gravity, whose Lagrangian density is of form, $\sqrt{-g} C_{\mu\nu\rho\sigma}^2$, 
is asymptotically free, and the issue of the perturbative unitarity is closely relevant to infrared dynamics of asymptotic fields,
so this problem becomes to be quite nontrivial. Provided that we can confine the ghosts in the Weyl theory
like in QCD, we would be free of the perturbative unitarity.

\end{document}